\documentclass[twocolumn,twoside,slac_two]{revtex4}
\usepackage{graphicx}
\usepackage{fancyhdr}
\def\simleq{\; \raise0.3ex\hbox{$<$\kern-0.75em \raise-1.1ex\hbox{$\sim$}}\; }
\def\simgeq{\; \raise0.3ex\hbox{$>$\kern-0.75em \raise-1.1ex\hbox{$\sim$}}\; }

\newcommand{\GeV}{{\rm GeV}}
\newcommand{\TeV}{{\rm TeV}}

\begin{document}

\title{The High Energy Cosmic Ray Electron Spectrum measured by
Fermi Gamma-Ray Space Telescope: some possible interpretations}

%
\author{G. Di Bernardo$^1$, D. Gaggero$^{1,2}$ and D. Grasso$^2$ \\ {\bf on behalf of Fermi-LAT collaboration}}
\affiliation{$^1$ Istituto Nazionale di Fisica Nucleare, Sezione di Pisa\\ 
$^2$ Dipartimento di Fisica dell'Universit\`a di Pisa, I-56127 Pisa, Italy}
%
%

\begin{abstract}
The Fermi Large Area Telescope (LAT) has provided the measurement of the high energy cosmic ray electrons plus positrons (CRE) spectrum with unprecedented accuracy form 20 GeV to 1 TeV. 
Recently this range has been extended down to  $\simeq 7~ \GeV$. The spectrum shows no prominent  features and it is significantly harder than that inferred from several previous experiments.  While the 
reported Fermi-LAT data alone may be interpreted in terms of a single (electron dominated) Galactic component, when combined with other complementary experimental results, specifically the CRE 
spectrum measured by H.E.S.S., and especially the positron fraction measured by PAMELA, an additional electron and positron component seems to be required. We show that the acceleration of 
electron-positron pairs in Galactic pulsars may offer a natural (though not unique) interpretation of all those results. 

\end{abstract}

\maketitle

\thispagestyle{fancy}


\section{Introduction}
Prior to 2008, the high energy electron spectrum was measured by balloon-born experiments \cite{Kobayashi:2003kp} and by a single space mission AMS-01 \cite{ams1}. Those data are compatible with a featureless power law spectrum within their errors. 
This is in agreement with theoretical predictions (for a recent review see \cite{Strong:2007nh}) assuming: i) that the source term of CR electrons is
treated as a time-independent and smooth function of the position in the Galaxy, and the energy dependence is assumed to be a power law; ii) that the propagation is described by a diffusion-loss equation whose effect is to steepen the spectral slope respect to the injection. 
Possible deviations from a simple power law spectrum may, however, be expected above several hundred GeV as a consequence of synchrotron radiation and 
Inverse Compton (IC) energy losses which, at those high energies, limit the electron propagation length to a distance comparable to the mean distance between 
astrophysical sources \cite{Aharonian:95,Pohl:1998ug} or because the possible presence of exotic sources.    

Almost a year ago, the ATIC balloon experiment \cite{atic} found a prominent spectral feature at around 600 GeV in the total electron spectrum.  Furthermore, the H.E.S.S. \cite{hess,hess:09} atmospheric Cherenkov telescope reported a significant steepening of the electron plus diffuse photon spectrum above 600 GeV. Another independent indication of the presence of a possible deviation from the standard picture came from the recent measurements of the positron to electron fraction, e$^+$ /(e$^-$+ e$^+$), between 1.5 and 100 GeV by the PAMELA satellite experiment \cite{PAMELA_Nature}.  PAMELA found that the positron fraction changes slope at around 10 GeV and begins to increase steadily up to 100 GeV. 
This behavior is very different from that predicted for secondary positrons produced in the collision of CR nuclides with the interstellar medium (ISM).

The experimental information available on the CRE spectrum has been drastically expanded as the Fermi Collaboration has reported a high precision measurement of the electron spectrum from 20 GeV to 1 TeV performed with its Large Area Telescope (LAT)  \cite{Fermi_CRE_1} based on the first 6 months of data taking. 
Preliminary updated Fermi-LAT measurements, based on the first 12 month data, basically confirm the high energy spectrum measured by Fermi above 20 GeV and extend it down to almost 7 GeV.  In that  low energy range the Fermi-LAT data are compatible with those of AMS-01. 
The electron + positron spectrum measured by Fermi-LAT is compatible with a simple power law of slope $\gamma = - 3.08$ and presents some hints of a hardening at around 70 GeV and  steepening above $\sim 500$ GeV.
Although the significance of those features is low within current systematics, they suggest the presence of more components in the electron high energy spectrum. 
It is also worth noticing here that the hard electron spectrum observed by this experiment exacerbates the discrepancy between the predictions of standard CR theoretical models and the positron faction excess measured, most conclusively, by PAMELA \cite{PAMELA_Nature}. 
This makes the exploration of some non-standard interpretations more compelling. 

\section{Conventional interpretation}\label{sec:GCRE}

In this section we discuss a simple possible interpretation of Fermi-LAT CRE data in terms of a conventional
{\footnote{by conventional we intend here models which are normalized to the locally observed CRE data}}   Galactic CRE
 scenario assuming that electrons sources are continuously distributed in the Galactic disk and that positrons are only produced by the collision of primary CR nuclides with the interstellar gas. To this purpose we use the GALPROP numerical CR propagation code \cite{Moskalenko:01}.
We start considering a conventional model assuming a single power-law dependence of the diffusion coefficient on rigidity with index $\delta = 0.33$ (Kolmogorov like)
and electron ($e^-$ only) injection spectral index $\gamma_0 = - 1.6/-2.42$ below/above 4 GeV.
In \cite{interpretation_paper} this model was shown to provide a good description of published Fermi-LAT data above 20 GeV  (few other combinations of $\delta$ and $\gamma_0$ were also shown to fit the data).  

\begin{figure*}[t]
\centering
\includegraphics[width=135mm]{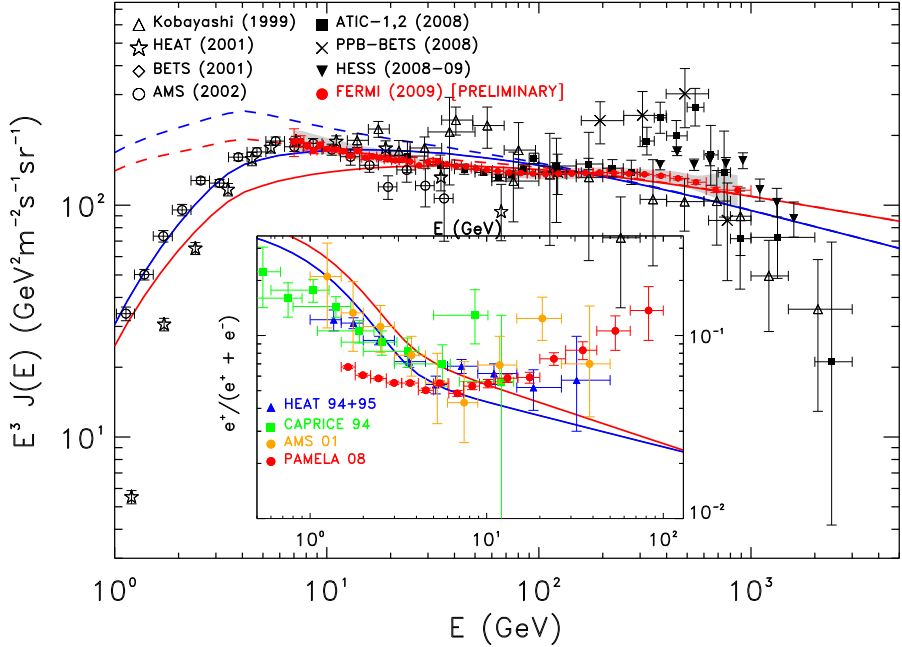}
\caption{ Fermi-LAT CRE data  \cite{Fermi_CRE_1}, as well as several other experimental data sets,
 are compared to the $e^- + e^+$ spectrum modeled with GALPROP (see \cite{Moskalenko:01}).
 The gray band represents systematic errors on the CRE spectrum measured by Fermi-LAT.
 The spectra are obtained with injection indexes $\gamma_0 = - 2.42/2.5$  (red/blue) above $4~\GeV$ and $\gamma_0 = - 1.6$ below
 that energy. The diffusion coefficient spectral slope is $\delta = 0.33$ for both models. In the insert  the positron fraction for the same models is compared with experimental data.
 Modulation in accounted in the force field approximation assuming a potential $\Phi = 0.45~{\rm GV}$. }
 \label{fig:elepos_242reac}
\end{figure*}

That model, however, faces a series of problems when compared with other experimental data and, most seriously, with the preliminary low energy Fermi-LAT, namely:
i) below 20 GeV that model predicts an excess of electrons + positrons respect to Fermi-LAT and AMS-01\cite{ams1} data when a reasonable solar modulation is considered
(see red line in Fig. \ref{fig:elepos_242reac});
ii) the model overshoot  H.E.S.S. data above 1 TeV; 
iii) the positron fraction  $e^+/(e^+ + e^-)$ they predict is not consistent with that measered by PAMELA \cite{PAMELA_Nature}
(see the insert in Fig. \ref{fig:elepos_242reac}).

While (ii) may be interpreted as a consequence of the stochastic nature of astrophysical sources (see Sec. 2.2 in \cite{interpretation_paper} and ref.s therein) or a high energy
cut-off in the source spectrum,  the other caveats are more serious and call for some revisions of the model.
 
Disregarding high energy ($E > 10~\GeV$) PAMELA data in this section, we have only to bother about the discrepancy of the conventional model with low energy data. 
It should be noted that below 10 GeV solar modulation is expected to affect significantly the spectra of electrons and positrons reaching the Earth. In the most simple scenario
modulation is treated as a charge independent cooling effect which can be quantified in terms of a parameter $\Phi$ having the dimensions of rigidity (force field approximation).  Such parameter is time dependent and it is generally tuned to reproduce CR proton data. In periods of low solar activity (as those in which Fermi, PAMELA and AMS-01 have been operating) its value typically ranges in the interval $400-600$ MV. None of those values, however, allows the conventional model considered in 
\cite{interpretation_paper} to perfectly match Fermi-LAT data (see red lines in Fig.\ref{fig:elepos_242reac}).
Assuming a steeper injection spectrum below few GeV's, while allows to improve the fit of Fermi-LAT data, it overshoots considerably AMS-01.
Such discrepancy may hard to be compensated even considering more complex charge dependent modulation set-up (see e.g. \cite{Potgieter}). 
Charge asymmetric modulation may also play a role explaining the discrepancy between the prediction of this model and low energy positron fraction PAMELA data \cite{Gast}.

A better fit can be obtained assuming a steeper spectrum at high energy, $\gamma_0 = - 2.5$ rather than $- 2.42$ which worsen the fit of high energy Fermi-LAT data (see blue line in Fig. \ref{fig:elepos_242reac}) but may still be compatible with the current statistical and systematic errors. Forthcoming Fermi-LAT data and analysis will, therefore, be crucial. 
Also this model, however, face serious problems when compared with low energy positron fraction PAMELA data.


A consistent interpretation of all available experimental data in the framework of single component conventional CR propagation models is therefore still challenging.  
 
\section{Extra component interpretation} 

Another viable interpretation of Fermi-LAT data consists in assuming the presence of two components in the CRE spectrum. This possibility was already risen in \cite{Fermi_CRE_1} and extensively discussed in \cite{interpretation_paper} and in a number of subsequent papers.
In this scenario, an extra electron and positron component of astrophysical (see e.g. \cite{Hooper:2008kg,Blasi:2009hv})
or exotic origin (see e.g. \cite{Bergstrom:2009fa,interpretation_paper,Meade:2009iu}) is added to the conventional one. The spectrum of the extra-component must be significantly harder that the conventional one and have an exponential cutoff at $E_{\rm cut} \simeq 1 ~\TeV$  such to match high energy Fermi-LAT and H.E.S.S. data, while the low energy background electron should be softer than in the single component scenario. It was showed that if the extra-component is composed of electrons and positrons in comparable amount it allows to explain the rising behaviour of the positron fraction observed by PAMELA above $10 ~\GeV$.   
In \cite{interpretation_paper} it was already noted as such scenario also allows to improve the fit of low energy pre-Fermi CRE data, especially those taken by AMS-01, respect to the single component conventional model.  In that paper an injection spectral index $-2.54$ was adopted for the Galactic $e^-$ conventional component. 

\begin{figure*}[ht]
\centering
\includegraphics[width=135mm]{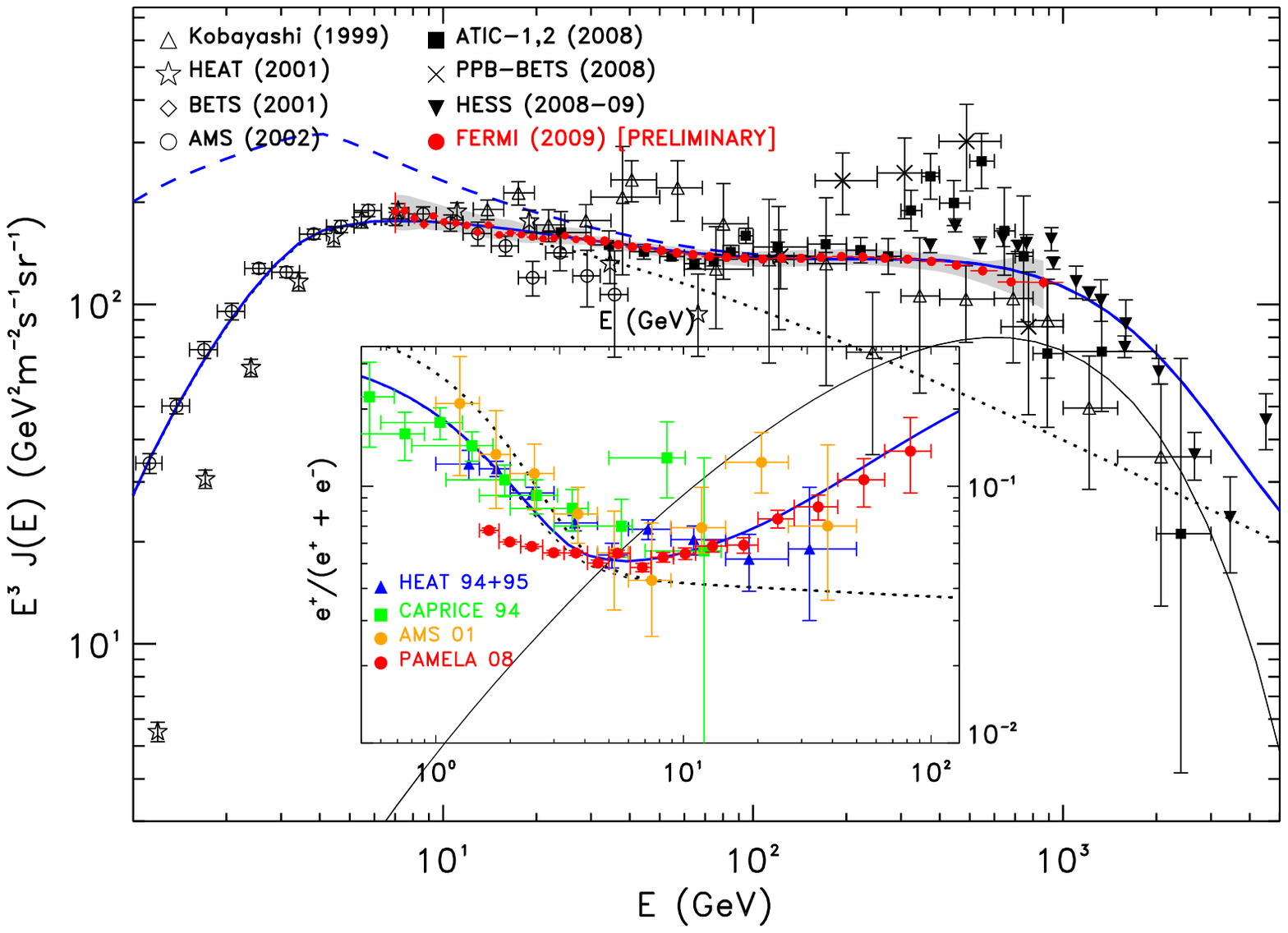}
\caption{This model assumes the presence of an electron and positron (charge symmetric) extra component with spectral index $\Gamma = -1.5$ and exponential cutoff at  $E_{\rm cut} \simeq 1.2 ~\TeV$ coming from a continuos distribution of sources in the Galactic disk (showed as a solid black line). The conventional electron component has an injection spectral index $\gamma_0 = -2.7$  ($\delta = 0.33$) (showed as a dotted black line). Both components are consistently propagated with GALPROP.  Solid lines are modulated in the force field approximation with $\Phi = 550~{\rm MV}$; dashed lines represent the LIS.  
 } 
 \label{fig:extra}
\end{figure*}

The new Fermi-LAT low energy data favour a steeper spectrum. As shown in Fig.\ref{fig:extra}, the combination of Galactic electron background with source spectral index $-2.7$, modulated in force-field approximation with $\Phi = 550~{\rm MV}$, and an electron + positron extra component with source spectral index $\Gamma = -1.5$  exponentially cut at 1 TeV, provides an excellent fit of the whole CRE spectrum measured by Fermi-LAT as well low energy AMS-01 data and H.E.S.S. data above 1 TeV.  PAMELA positron fraction is also nicely reproduced above 10 GeV while a small discrepancy remains  below that energy (see insert in Fig.\ref{fig:extra}). In that figure the extra component source distribution has been assumed to coincide with that of Galactic pulsars. We checked, however, that Fermi and PAMELA data can also be reproduced with a source distribution which resemble that expected from dark-matter annihilation (though with a slightly different choice of the relevant parameters).

A much better fit of low energy PAMELA data is obtained if a larger value of the diffusion coefficient spectral slope $\delta$ is adopted.
For example we found that if $\delta \simeq 0.5$ (Kraichnan like diffusion) as suggested by a recent analysis \cite{DiBernardo:2009ku} of CREAM B/C measurements  \cite{CREAM} and $\gamma_0 \simeq - 2.6$
the whole PAMELA and Fermi-LAT data set can consistently be fitted (the figures corresponding to that model will be showed elsewhere).

\begin{figure*}[ht]
\centering
\includegraphics[width=135mm]{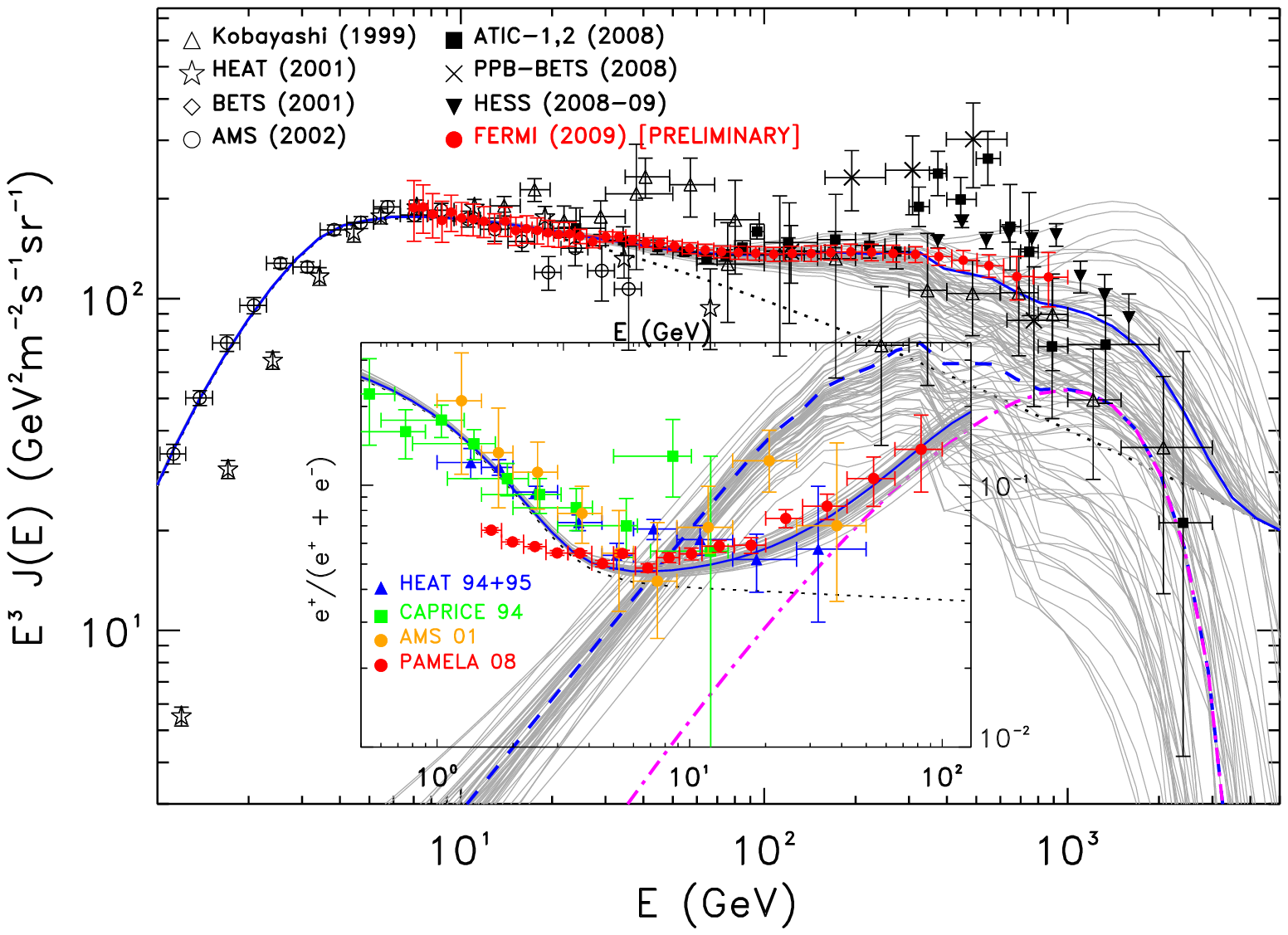}
\caption{The $e^- + e^+$ spectrum from pulsars (gray bottom lines) plus the Galactic conventional component (dotted line) is compared with experimental 
  data. Each gray top line represents the sum of all pulsars for a particular combination of pulsar parameters.  
  The dashed (pulsars only) and solid (pulsars + GCRE component) blue lines
   correspond to a representative choice among that set of possible realizations.  
  The dot-dashed (purple) line represents the contribution of Monogem pulsar in that particular case.
  Note that merely for graphical reasons, here Fermi-LAT statistical and systematic errors are added in quadrature.
  In the insert the positron fraction for the same models is compared with experimental data.
  Solar modulation is accounted as done in Fig.\ref{fig:extra}.}
   \label{fig:pulsar_model}
\end{figure*}

As a specific realisation of such scenario we considered a model in which the extra electron and positron component is produced by galactic pulsars as illustrated in details in \cite{interpretation_paper}.  Observed pulsars within 3 kpc from the Earth, as extracted from the ATNF radio pulsar catalogue (http://www.atnf.csiro.au/research/pulsar/psrcat/ ), 
are accounted. We tested that adding more distant pulsars have negligible effects on our results. 
For each of those pulsars we use the spin-down luminosity given in the catalogue and randomly vary the relevant parameter in the following representative ranges:
$1 < E_{\rm cut} < 1.8~\TeV$,  $10 < \eta_{e^\pm} <  40~\%$ ($e^\pm$ production efficiency),  $5 <  (\Delta t / 10^4~{\rm yr} ) <  10$ (delay of $e^\pm$ release respect to the pulsar birth)
and $1.5 < \Gamma < 2.0$. These ranges of parameter are compatible with our observational and theoretical knowledge of particle acceleration in PWNe.
The electron+positron spectra expected for any of those realizations, as well as the predicted positron fraction, are shown in Fig.\ref{fig:pulsar_model} in the case $\delta = 0.33$. 
It evident that are several realizations allowing to match all electron data as well as high energy PAMELA positron fraction data.

\section{Conclusions}\label{sec:conclusions}

We reported on possible interpretations for the cosmic ray electron-plus-positron (CRE) spectrum measured by Fermi-LAT including preliminary new data extending down to $E \simeq 7~ \GeV$ \cite{Luca,Melissa}.
We discussed the case of a single Galactic electron component, and a two-component scenario which adds to the conventional electron flux a primary electron and positron component with a harder spectrum, which may be either of astrophysical or exotic origin.  

In the single component scenario, Fermi-LAT electron data can be approximatively be described if an injection spectral index between $- 2.4$ and $- 2.5$ is assumed (for $\delta = 0.33$). 
Those models, however,  do not allow a perfect match of Fermi-LAT data either at low or at high energy. Furthermore they are in sharp tension with the PAMELA data on the positron fraction. 

We therefore considered models adding an extra electron and positron component to the conventional electron Galactic sea. We showed that such models allow to improve considerably 
the fit of Fermi-LAT, H.E.S.S. and AMS-01 electon data as well as to account for the rising behaviour of the positron fraction observed, most conclusively, by PAMELA.

\bigskip
\begin{acknowledgments}
The Fermi LAT Collaboration acknowledges support from a number of agencies and institutes for both development and the operation of the LAT as well as scientific data analysis. These include the NASA and the DOE United States, CEA/Irfu and IN2P3/CNRS in France, ASI and INFN in Italy, MEXT,
and the K. A. Wallenberg Foundation, the Swedish Research Council and the Swedish National Space Board in Sweden. Additional support from 
INAF in Italy for science analysis during the operations phase is also gratefully acknowledged. 
D.G. is supported by the Italian Space Agency under the contract AMS-02.ASI/AMS-02 n.I/035/07/0 and partially by UniverseNet EU Network under contract n. MRTN-CT-2006-035863.  

\end{acknowledgments}

\bigskip 

\end{document}